\documentclass[aps,pre,twocolumn,showpacs]{revtex4}
\usepackage{graphicx}
\begin{document}
\title{ Lattice stretching bistability and dynamic heterogeneity
}
\author{P. L. Christiansen}
\affiliation{Department of Informatics and Department of Physics, Technical
University of Denmark, DK 2800, Kgs. Lyngby, Denmark}

\author{A. V. Savin}
\affiliation{Semenov Institute of Chemical Physics, Russian Academy
of Sciences, Moscow 119991, Russia}

\author{A. V. Zolotaryuk}
\affiliation{Bogolyubov Institute for Theoretical Physics, National Academy of
Sciences of Ukraine, Kyiv 03680, Ukraine}

\date{\today}

\begin{abstract}
A simple one-dimensional lattice model is suggested to describe the experimentally observed 
plateau in force-stretching diagrams for some macromolecules. This chain model involves the 
nearest-neighbor interaction of a Morse-like potential (required to have a saturation branch) 
and an harmonic second-neighbor coupling. Under an external stretching applied to the chain ends,
the intersite Morse-like potential results in the appearance of a double-well potential within 
each chain monomer, whereas the interaction between the second neighbors provides a homogeneous 
bistable (degenerate) ground state, at least within a certain part of the chain. 
As a result, different conformational changes occur in the chain under the external forcing. 
The transition regions between these conformations are described as topological solitons. 
With a strong second-neighbor interaction, the solitons describe the transition between the 
bistable ground states. However, the key point of the model is the 
appearance of a heterogenous structure, when the second-neighbor coupling is sufficiently weak. 
In this case, a part of the chain has short bonds with a single-well potential, whereas 
the complementary part admits strongly stretched bonds with a double-well potential. This case allows 
us to explain the existence of a plateau in the force-stretching diagram for DNA and 
alpha-helix protein. Finally, the soliton dynamics are studied in detail. 
\end{abstract}

\maketitle

\section{Introduction}

\begin{figure}[tbp]
\begin{center}
\includegraphics[angle=0, width=1\linewidth]{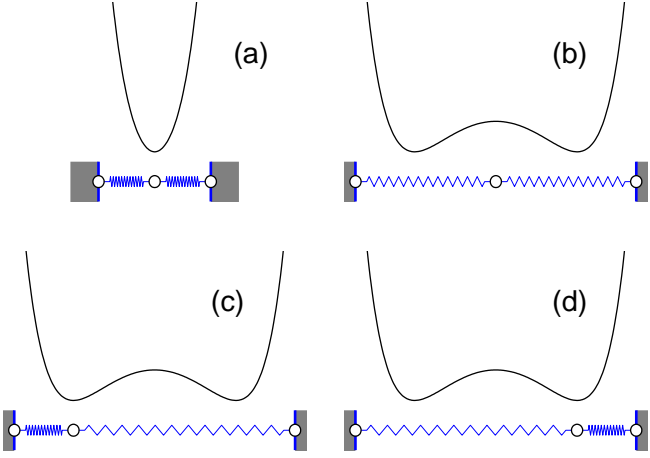}
\end{center}
\caption{\label{fig1}\protect\small (Color online)
(a) Single-well and (b-d) double-well potentials for a central particle interacting with its
fixed neighbors via inter-atomic Morse-like potential $V(r)$ and formed as the sum $V(r)+V(2R-r)$. 
(a) Distance $2R$ between the lateral fixed particles does not exceed $2R_0$, where $R_0 $
is a point of inflection of $V(r)$. (b) Unstable equilibrium position of the central particle. 
(c) Left and (d) right stable equilibria of the central particle.
}
\end{figure}

The effect of appearance of bistable states caused by lattice stretching has been first studied 
by Manevitch {\it et al.} \cite{s0} for modeling the mechanodestruction of a polymer chain. 
As a simple model, the anharmonic chain with the nearest-neighbor coupling in the form 
of the Morse potential has been chosen. Under lengthening this chain by applying an external force 
to its ends, the formation of an effective double-well potential in the chain bonds has been shown.
The idea of this conformational transition can be explained in simple terms as follows. 
Consider three coupled particles as shown in Fig.~\ref{fig1}, where the two lateral particles are 
fixed and the central particle interacts with its neighbors through a Morse-like potential $V(r)$ 
with a minimum at $R=r_0$ and a constant asymptote 
$\varepsilon = \lim_{r \to \infty}V(r)$. 
The potential of this type has a point of inflection at $r=R_0 >r_0$. The Morse potential 
\begin{equation}
V(r) = \varepsilon \left[ {\rm e}^{-\beta
(r -r_0)} -1 \right]^2 , ~~ 0< r < \infty , 
\label{f1}
\end{equation} 
given in dimensionless units, 
with any parameter $\beta > 0$, can be chosen as a particular example
where $R_0=r_0+\beta^{-1}\ln 2$. The total potential for the middle particle is 
$V(r)+V(2R-r)$, where $2R$ is the distance between the lateral particles.
Using the new variable $u=R-r$, this potential can be written in a more convenient 
symmetric form as follows 
\begin{eqnarray}
&&W(u)\doteq  V(R-u)+V(R+u)= \varepsilon e^{-2\beta(R-r_0)} \nonumber \\
&& \times \{[2\cosh(\beta u)-e^{\beta(R-r_0)}]^2
+e^{2\beta(R-r_0)}-2\}. \label{1a}
\end{eqnarray} 
The potential $W(u)$ has only one minimum $u=0$ if $R\le R_0$ 
as demonstrated by Fig.~\ref{fig1}(a), and one maximum $u=0$ and the two minima
$u=\pm u_0$ if $R>R_0$ [see Fig.~\ref{fig1}(b-d)]. For the potential (\ref{f1}),
$u_0(R)=\beta^{-1}\mbox{arccosh}(e^{\beta(R-r_0)}/2)>0$, being the solution of the equation 
\begin{equation}
\cosh(\beta u_0)={\rm e}^{\beta (R-r_0)}/2.
\label{f5}
\end{equation}

The three-particle system illustrated by Fig.~\ref{fig1} can be extended to a finite chain 
consisting of $2N+1$ particles (or $N$ monomers), where the terminal particles are fixed and 
the total chain length is $L=2NR$. If $R>R_0$, double-well potentials can be formed inside the chain.
In this case, many ground states of the chain are possible resulting in different irregular  chain conformations.
Therefore the model studied in Ref.~\cite{s0} has to be modified in such a way that a sufficiently stretched
chain would admit homogeneous ground states with periodic structure. To this end, we involve additionally
a stabilizing second-neighbor interaction and, as a result, the ground states with alternating lengths of chain bonds are possible.  

\begin{figure}[tbp]
\begin{center}
\includegraphics[angle=0, width=1\linewidth]{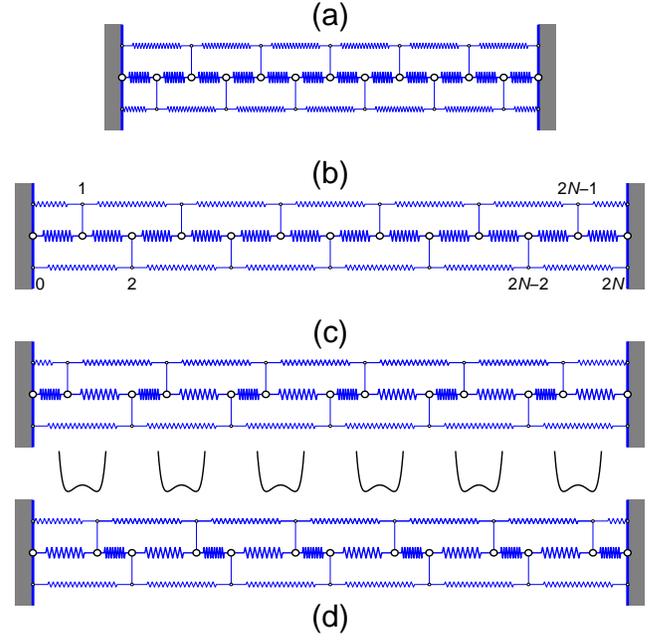}
\end{center}
\caption{\label{fig2}\protect\small (Color online)
Equilibria in monoatomic chain with fixed terminal particles, where except for the nearest-neighbor
interactions, also a coupling between the second neighbors is involved: (a) Stable (ground) state 
of the stretched chain with period $R\le R_0$, where each atom is found in a single-well potential.
(b) Unstable state of the chain, where all atoms with odd number are located at the top of a double-well potential.
(c) Left and (d) right degenerate ground states of the stretched chain.
}
\end{figure}

It is sufficient to impose an harmonic coupling between the second neighbors as shown schematically
in Fig.~\ref{fig2}. Let $K$ be a (dimensionless) stiffness constant of the second-neighbor interaction
with $x_0, x_1, \ldots, x_{2N}$ being  positions of the chain atoms.
Then the total potential energy of the $N$-monomer system with fixed terminal atoms can be written as
\begin{eqnarray}
 E_{N} &=& {1\over 2} \sum_{n =0}^{2N-1}  V(x_{n+1}-x_n) 
+  K (x_{1}-x_0 - r_0)^2 \nonumber \\ 
&& + \sum_{n =0}^{2N-2} {1 \over 2} K (x_{n+2}-x_n - 2r_0)^2 
\nonumber \\ 
&& + \, {1 \over 2} K (x_{2N}-x_{2N-1} - r_0)^2. 
\label{f2} 
\end{eqnarray}

It is expected that a sufficiently strong stretching of the chain results in a dimeralization 
of the chain, for which the even atoms are found in equilibria
$x_n = nR$ with $n= 0, 2, \ldots , 2N$. Inserting these values into the energy (\ref{f2}), we find
\begin{eqnarray}
 E_{N} =N \left[ V(r) + V(2R-r) \right] 
 +4(N-1)K(R-r_0)^2\nonumber\\
 +\frac12K(r-r_0)^2+\frac12K(2R-r-r_0)^2.
\label{f3}
\end{eqnarray}
Differentiating Eq.~(\ref{f3}) with respect to $r$, we find the following equation for equilibria:
\begin{equation}
  V'(r) - V'(2R-r) = (2K/N )(R-r).
\label{f4}
\end{equation}
Using the variable $u=R-r$, Eq.~(\ref{f4}) becomes $W'(u)=-2Ku/N$.
The trivial solution $u=0$ describes equilibria of 
odd atoms (stable if $R\le R_0$ and unstable if $R>R_0$). The two stable equilibria with $u=\pm u_0$ appear when $R>R_0$ and $N\gg 1$.
For a long chain ($N\to\infty$), the role of boundary conditions can be neglected, so that 
Eq.~(\ref{f4}) for the equilibrium positions $\pm u_0$ in the case of the Morse potential (\ref{f1})
takes the form of Eq.~(\ref{f5}), so that the two stable minima exist if the inequality $R>R_0$ is fulfilled.

To simplify the problem with fixed chain ends, it is convenient to
use the cyclic boundary conditions by putting 
\begin{equation}
x_{2N}=x_0+2NR~~~\mbox{and}~~~x_{2N+1}=x_1+2NR. 
\label{6a}
\end{equation}
Then, if we fix the whole chain by setting $x_0\equiv 0$, 
so that $x_1=\pm u_0$, the equilibrium positions are 
\begin{equation}
 x_{n}^0 = \left\{ \begin{array}{ll}
  nR ~~~~~~~~~ \mbox{if} ~~ n= 0, 2 , \ldots , 2N , \\
  nR \pm u_0 ~~ \mbox{if} ~~ n= 1, 3 , \ldots , 2N +1 ,
  \end{array} \right.
 \label{f6}
\end{equation}
The subscripts $``+ "$ and ``-" denote the two degenerate ground states in the 
dimeralized chain, respectively. 
Schematically, these states can be represented as\linebreak
 $X$--$X$------$X$--$X$------$X~~\cdots~~X$--$X$------$X$--$X$------$X~~~$ and
 $X$------$X$--$X$------$X$--$X~~\cdots~~X$------$X$--$X$------$X$--$X$,
where the terminal $X$'s are fixed and all the bulk atoms are found either in the left well 
or the right well, respectively. Obviously, the domain walls
(topological kinks and antikinks) that separate these two ground states can be excited.
However, this is true if the second-neighbor interaction is sufficiently strong, at least for the model suggested in this paper. As shown below, the situation appears more complicated for a weak second-neighbor coupling, the case being of experimental relevance for some macromolecules. More precisely, the existence of a plateau in the force-stretching diagrams for DNA double helix 
\cite{d1,d2,d3,d4,d5} as well as for $\alpha$-helices of protein \cite{d6} can be explained within the framework of our model.

The paper is organized as follows. In the next section, we present the equations of motion for a stretched
nonlinear monoatomic chain and discuss the spectrum of small-amplitude oscillations.  The analysis 
of switching a bistable ground state of the chain is given in Sec.~III. In Sec.~IV, we find the profiles of kink and antikink solutions and study their dynamical properties. 
The next section is devoted to realistic systems, where the topological soliton solutions obtained in the previous section are studied in detail.
Conclusions are given in Sec.~VI.

\section {A model and its linearized version}

With the notations introduced in the previous section, the (dimensionless) Hamiltonian for the monoatomic chain model
with the cyclic boundary conditions (\ref{6a}) can be written in the form
\begin{eqnarray}
 H =\sum_{n =0}^{2N-1} \left[ {1 \over 2} m \dot{x}_n^2
 +  V(x_{n+1}-x_n)\right. \nonumber \\
 \left.
 + \, {1 \over 2}K(x_{n+2}-x_n - 2r_0)^2 \right] , \label{f7}
\end{eqnarray}
where $m$ is a chain particle mass and the dot denotes the differentiation over time $t$.
Here the strings connecting the second neighbors are assumed to be undistorted at the length $2r_0$. The corresponding equations of motion are
\begin{eqnarray}
m\ddot{x}_n =
  \left[ V'(x_{n+1}-x_n) -V'(x_n -x_{n-1}) \right]\nonumber\\
 +  K(x_{n+2}-2x_n + x_{n-2}) ,
\label{f8}
\end{eqnarray}
where $n=0,1, \ldots , 2N-1$.
\begin{figure}[tbp]
\begin{center}
\includegraphics[angle=0, width=1\linewidth]{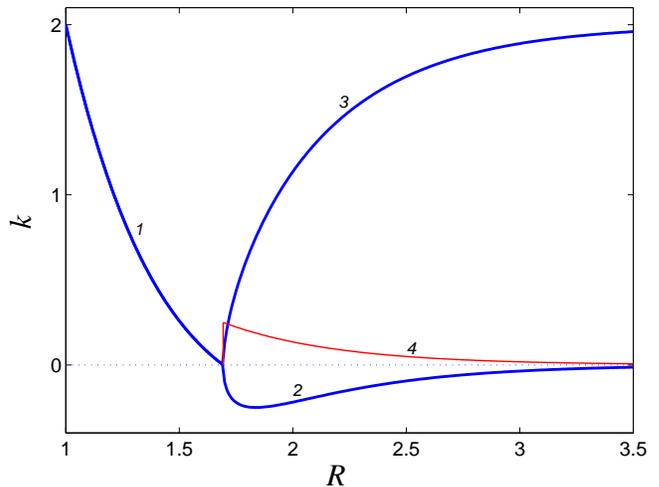}
\end{center}
\caption{\label{fig3}\protect\small (Color online)
Stiffness $k=V''(R \pm u_0)$ as a function of lattice constant $R$ calculated for potential 
(\ref{f2}) with $\varepsilon=1$, $\beta=1$, and $r_0=1$: $k=k_\pm$ for 
$R\le R_0$ (curve 1), $k=k_+$ (curve 2) and $k=k_-$ (curve 3) for $R>R_0$. 
 Critical stiffness $k=k_c=-k_-k_+/2(k_-+ k_+)$ is given by curve 4.
}
\end{figure}

The linearized version of Eqs.~(\ref{f8}) is obtained by putting
$x_n(t)=x_{n}^0+v_{n}(t),~n=0,1,\ldots,2N+1$, where the equilibria $x_n^0$'s are defined by Eqs.~(\ref{f6}). As a result, we find 
\begin{eqnarray}
m\ddot{v}_n = \left[ k_\pm (v_{n+1}-v_n)-k_\mp (v_{n}-v_{n-1})\right]
\nonumber \\
 +  K(v_{n+2}-2v_n + v_{n-2}),
\label{f9}
\end{eqnarray}
where $k_\pm = V''(R \pm u_0)$. The upper subscript at the
stiffness coefficient belongs to the particles with even
(odd) $n$'s and the lower one to those with odd (even) $n$'s for the case when all the odd particles are found in the right (left) well of the double-well potential.

Under stretching $R\le R_0$, the displacement $u_0$ becomes zero and therefore we have the 
stiffness constant $k=k_-=k_+ =V''(R)$.  With the stretching of the chain, the stiffness 
$k(R)$ decreases monotonically and at $R=R_0$ it reaches zero (Fig.~\ref{fig3}, curve 1). At further 
lengthening, the chain becomes bistable and the distance $2u_0$ increases monotonically with the 
growth of $R$. The stiffness of the short bonds $k_-=\varepsilon V''(R-u_0)>0$ increases monotonically
(Fig.~\ref{fig3}, curve 3) and the stiffness of the long bonds becomes negative: $k_+=V''(R+u_0)<0$
(curve 2), but their sum is always positive: $k_-+k_+>0$. Explicitly, for the potential (\ref{f1}) we get
\begin{equation}
k_\pm=\mp\varepsilon\beta^2{\rm e}^{\mp\beta u_0}\sinh(\beta u_0)\cosh^{-2}(\beta u_0),
\label{f10}
\end{equation}
so that in this case $k_-+k_+ =2\varepsilon \beta^2 \tanh^2(\beta u_0) >0$.
\begin{figure}[tbp]
\begin{center}
\includegraphics[angle=0, width=1\linewidth]{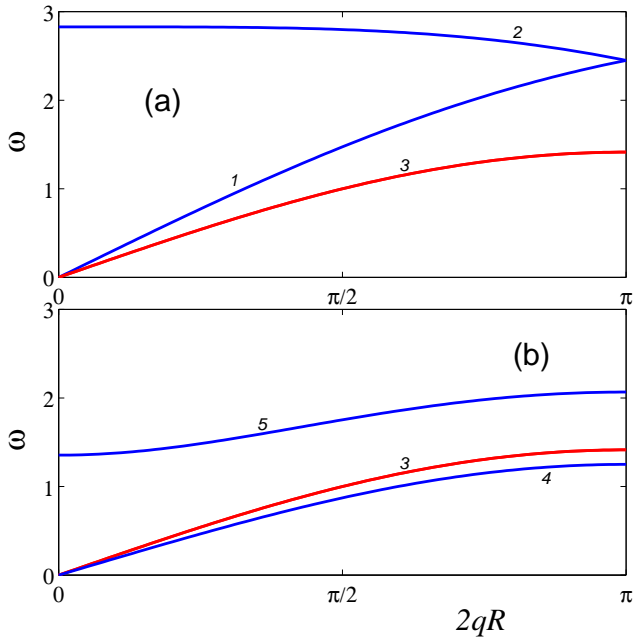}
\end{center}
\caption{\label{fig4}\protect\small (Color online)
Linear spectra $\omega =\omega_\pm(2qR)$ calculated (at $m=1$) in the case of potential 
(\ref{f1}) with 
$\varepsilon=1$, $\beta =1$, $r_0=1$, and $K=0.5$ for (a) $R=r_0$ (curves 1 and 2), $R=R_0$ (curve 3),
and (b) $R=2>R_0$ (curves 4 and 5). Curves 1,3, and 4 relate to 
$\omega_-(2qR)$ and curves 2,3, and 5 relate to $\omega_+(2qR)$. 
}
\end{figure}

The dispersion law is obtained if we insert the small-amplitude waves
\begin{equation}
 v_n(t) =  A {\rm e }^{{\rm i}( qnR -\omega t) }
 ~~\mbox{and}~~
  v_{n+1}(t) =  B {\rm e }^{{\rm i}( qnR -\omega t) }
\label{f11}
\end{equation}
into Eqs.~(\ref{f9}) with $n=0, 2, \ldots , 2N$. Here $\omega$ is the frequency and the wave number $q \in [0,  \pi/2R)$. As a result, we find the
dispersion law that admits two branches of the spectrum (acoustic and optical):
\begin{eqnarray}
m\omega_{\pm}^2 &=& k_- + k_+  + 2K[1-\cos(2qR)]\nonumber \\
&&\pm \sqrt{k_-^2 +k_+^2 + 2 k_- k_+ \cos(2qR)  } \, .
\label{f12}
\end{eqnarray}

The dispersion curves for $R\le R_0$ are present in Fig.~\ref{fig4}(a). For $R<R_0$ the curve 
$\omega_-(q)$ is the continuation of the curve $\omega_+(q)$. They can be considered as a 
single acoustic branch. In the limit $R\rightarrow R_0-0$, the curves $\omega_-(q)$ and 
$\omega_+(q)$ merge 
together into a single curve $\omega^2=2K [1-\cos(2qR)]$. At further increase of $R$, 
this curve splits into two disconnected curves $\omega_-(q)$ and $\omega_+(q)$ 
[Fig.~\ref{fig4}(b), curves 4 and 5], where the first curve corresponds to acoustic oscillations, while the second one to optical oscillations of the stretched chain.

At stretching $R<R_0$, the uniformly stretched state is always stable, since all the coupling 
constants $k_-$, $k_+$, and $K$ are positive. The situation changes when $R>R_0$ because here one of the coupling constants,
$k_+$, is negative and now the stability of the alternating state of the stretched chain depends on the stiffness constant $K$. For the stability it 
is necessary that the inequality
\begin{eqnarray}
k_- + k_+  + 2K [1-\cos(2qR)]\nonumber \\
-\sqrt{k_-^2 +k_+^2 + 2 k_- k_+  \cos(2qR) }>0 
\label{f13}
\end{eqnarray}
has to be fulfilled for all values of the wave number $q>0$. It is easy to show that this 
condition holds only if
\begin{equation}
K> k_c (R) \doteq -k_-k_+/2(k_-+k_+). 
\label{f14}
\end{equation}

The dependence of the critical value of the stiffness of the second-neighbor interaction 
 on $R$ is given in Fig.~\ref{fig3} (curve 4). For all $R$ this value is positive; it 
monotonically decreases with increase of the chain stretching $R$. Its maximum reaches in the limit $R\rightarrow R_0+0$ when $u\to u_0 +0$. Using Eq.~(\ref{f10}), 
for the particular case (\ref{f1}) we obtain
\begin{equation}
k_c(R) = {\varepsilon \beta^2 \over 4} \cosh^{-2}(\beta u_0),
\label{f15}
\end{equation}
where the dependence of $u_0$ on $R$ is given by Eq.~(\ref{f5}). Next, we find the limiting value $ K_0 \doteq \lim_{R\rightarrow R_0+0}k_c(R)=\varepsilon\beta^2/4 $ and therefore 
the stability condition of the 
alternating states of the stretched chain takes the following simple form:
\begin{equation}
K> K_0 .
\label{f17}
\end{equation}
Thus, for the stability of the stretched alternating states of the chain, it is necessary and sufficient 
that the stiffness of the interaction of the second neighbors has to be greater the eighth part of the stiffness of the interaction the nearest neighbors of the unstretched chain.

The velocity of long-wave acoustic phonons 
$v_0=\lim_{q\rightarrow 0}\omega(q)/q$ can be calculated
from the spectra (\ref{f12}) for different values of $R$. As a result, we obtain
\begin{equation}
v_0=\left\{ \begin{array}{ll} 2R\sqrt{(k/4 +K)/m}~~~\mbox{if}
~~R\le R_0, \\
2R\sqrt{( K - k_c )/m }~~~~~\mbox{if}~~R > R_0. 
\end{array} \right.
\label{f18}
\end{equation}
Note that the condition (\ref{f14}) ensures the positivity of the expression under the radicals in (\ref{f18}).

\section{Transition to the bistability of the ground state under stretching the chain}

In order to understand how the ground state of the chain changes under its stretching, we consider 
the dependence of the ground energy $E$ of the homogeneous chain state on the lattice spacing $R$. 
For the uniformly stretched chain state, when  $x_{n+1}-x_n=R $, $ 0 < R < \infty$, $n=0, 1,\ldots , 2N$, the deformation energy of one chain unit is
\begin{eqnarray}
E_1(R)&=& V(R)+2K(R-r_0)^2\nonumber\\
&=&\varepsilon[e^{-\beta(R-r_0)}-1]^2+2K(R-r_0)^2. \label{18a}
\end{eqnarray}
On the other hand, when we consider the ground state with the alternating bond lengths 
$R-u_0$ and $R+u_0$ ($R > R_0$), the deformation energy of one chain unit becomes
\begin{eqnarray}
E_2(R)&=&W(u_0)/2+2K(R-r_0)^2\nonumber\\
&=& {\varepsilon \over 2} [1-2e^{-2\beta(R-r_0)}]+2K(R-r_0)^2. \label{18b}
\end{eqnarray}

\begin{figure}[tbp]
\begin{center}
\includegraphics[angle=0, width=1\linewidth]{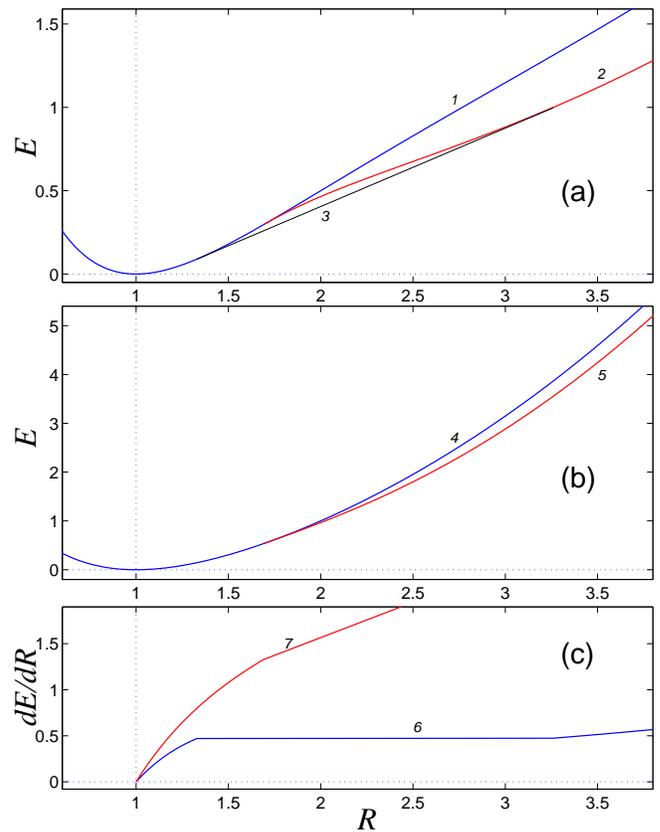}
\end{center}
\caption{\label{fig5}\protect\small (Color online)
Dependence of energy $E$ and its derivative $dE/dR$ of uniformly stretched state of the chain with equal bonds $E=E_1$ (curves 1 and 4) and that with alternating bonds $E=E_2$ 
(curves 2 and 5) on lattice spacing $R$ for (a) and (c), line 6, $K=0.05$; and (b) and (c), line 7, $K=0.3$ ($\beta=1$, $\varepsilon=1$, $r_0=1$, $K_0=0.25$, and $R_0 = 1 + \ln 2 = 1.693$). Lines 3 and 6 give convex neighborhood of function $E(R)$.
}
\end{figure}

For comparison the form of the functions $E_1(R)$ and $E_2(R)$ is depicted in Fig.~\ref{fig5}.
The function $E_1(R)$ has a minimum at $R=r_0$, increasing for $R > r_0 $. At $R=R_0$ both these functions are smoothly ``sewed'' together because
$E_1(R_0)=E_2(R_0)=\varepsilon/4 + 2K (R_0 -r_0)^2$ and $E_1'(R_0)=E_2'(R_0)=\varepsilon\beta/2+4K(R_0-r_0)$. However, for $R>R_0$ the function 
$E_2(R)$ steps aside smoothly and continues further below  
$E_1(R)$. Therefore the energy of the homogeneously stretched (with any $R$)
ground state of the chain is given by the smoothly sewed function 
$E(R)=E_1(R)$ for $R\le R_0$ and $E(R)=E_2(R)$ for $R\ge R_0$. At $R=R_0$ the second derivative 
$E_2''(R_0)=4K-\varepsilon\beta^2$ is positive if $K>K_0$ 
and negative if $K<K_0$. Therefore $E(R)$ is a strongly concave function (for all $R$) only if
$K>K_0$ [see Fig.~\ref{fig13}(b)]. In this case, the ground state of the chain is always the 
homogeneous conformation with equal bond lengths for  $R\le R_0$ and that with alternating bonds 
for  $R>R_0$. The inequality  (\ref{f17}) ensures the stability of the uniformly stretched 
state of the chain. 

For $K<K_0$ a local convexity in the $E(R)$ behavior, as illustrated in Fig.~\ref{fig5}(a) by 
curve 2, appears in a neighborhood of $R_0$, i.e., on some interval $R_1<R<R_2$ with $R_1 < R_0$ 
and $R_2 > R_0$. This means that the homogeneous state given by the energy $E_2(R)$, $R_1 < R< R_2$,
with the alternating bond lengths  $R-u_0(R)$ and $R+u_0(R)$ in fact is unstable. Instead, 
a {\it heterogeneous} conformation, where some part of the chain has equal bonds and the other one alternating bonds, appears more stable. In this case, the chain energy behavior 
can be obtained by connecting the two points 
$\{R_1,E(R_1)\}$ and $\{R_2,E(R_2)\}$ by a line [see Fig.~\ref{fig5}(a), line 3]. In other words, moving
along this line, the heterogeneous state with one part of the chain being in a weakly stretched state
with equal bonds and the spacing $R_1$, and the other part in a strongly stretched state with alternating bond lengths $R_2-u_0(R_2)$, $R_2+u_0(R_2)$ and the spacing $R_2$
appears more energetically favorable. In this case, the heterogeneous stretching (lengthening) of the
whole chain occurs due to the increase of the portion of strongly stretched bonds. This scenario of
the heterogeneous stretching results in the appearance of the stationary region (plateau) in the force-stretching diagram under the chain lengthening ($R_1 < R < R_2$) as illustrated by line 6 in Fig.~\ref{fig5}(c).  

\section{Dynamic heterogeneity and Topological solitons}

Since the nonlinear lattice model introduced in the previous section admits the heterogeneous 
structure that appears to be energetically favorable, the existence of freely moving topological defects is expected.  The corresponding soliton solutions can be 
found numerically using the steepest-descent method. To this end, it is convenient to use the variables: coordinates $u_n=x_n/r_0$, energy ${\cal H}=H/\varepsilon$, and time 
$\tau= r_0^{-1}\sqrt{\varepsilon/m} \, t$. Then the dimensionless Hamiltonian of the cyclic chain takes the form 
\begin{eqnarray}
{\cal H}=\sum_{n =0}^{2N-1} \left[ {1 \over 2} 
\left(d u_n \over d\tau\right)^2
 + {\cal V} (u_{n+1}-u_n)\right. \nonumber \\
 \left.  + {1 \over 2}\kappa (u_{n+2}-u_n - 2)^2 \right], 
\label{f19}
\end{eqnarray}
where ${\cal V}(u_{n+1} -u_n) =\varepsilon^{-1} V(x_{n+1}-x_n)$ and 
$\kappa = Kr_0^2/\varepsilon$. According to Eqs.~(\ref{6a}), the chain tension is given through the boundary conditions:
\begin{equation}
u_{2N}\equiv u_0+2Na~~~\mbox{and}~~~u_{2N+1}\equiv u_1+2Na, 
\label{f20}
\end{equation}
where the lattice spacing of the stretched chain is $a = R/r_0\ge 1$.

The system of equations
\begin{eqnarray}
{d^2 u_n \over d\tau^2 } &=& {\cal V}'(u_{n+1}-u_n)-{\cal V}'(u_n-u_{n-1})\nonumber \\
&&+ \, \kappa (u_{n+2}-2u_n+u_{n-2}),  \label{f21}
\end{eqnarray}
where $n=0,1, \ldots ,2N-1$, corresponds to the Hamiltonian function (\ref{f19}) with the boundary conditions (\ref{f20}) and $ u_{-1}\equiv u_{2N-1}-2Na$ and 
$u_{-2}\equiv u_{2N-2}-2Na$. 
For the relative displacements $r_n=u_{n+1}-u_n$, the equations of motion 
(\ref{f21}) become
\begin{eqnarray}
{d^2 r_n \over d\tau^2} &=& {\cal V}'(r_{n+1})-2{\cal V}'(r_n)+{\cal V}'(r_{n-1}))
\nonumber \\
&&+ \, \kappa (r_{n+2}-2r_n+r_{n-2}).  \label{f23} 
 \end{eqnarray}

For numerics we choose the following values of the parameters: $\varepsilon=1$,
$\beta=1$, and $r_0=1$. Then for $\kappa >\kappa_0=1/4$, the ground state will always 
be uniformly stretched because the stability condition of the alternating states 
(\ref{f17}) is fulfilled. Consider the chain with the lattice 
spacing $a>a_0=R_0/r_0=1+(\beta r_0)^{-1}\ln 2$ and $\kappa>\kappa_0$. Then the chain has the 
following two ground states with equal energy: $r_{2n-1}=a\mp\delta$ and 
$r_{2n}=a\pm\delta,~n=1, \ldots ,N/2$, where $\delta = u_0/r_0 $. We look for traveling wave 
solutions of the equations of motion (\ref{f23}) that describe the smooth transition of the chain from one ground state to the other one, i.e., we put
\begin{equation}
r_{2n-1}(\tau)=r_1(2na-s\tau),~~r_{2n}(\tau)=r_2(2na-s\tau) , 
\label{f24}
\end{equation}
where $s$ is a dimensionless traveling wave velocity.
Suppose next that the form of the traveling wave (\ref{f24}) smoothly depends on the lattice 
number $n$. Then the second derivatives over time can approximately be substituted by the discrete derivatives as follows:
\begin{eqnarray}
{d^2 \over d\tau^2} r_{2n-1}& \simeq &s^2(r_{2n-3}-2r_{2n-1}+r_{2n+1})/4a^2, \nonumber \\
{d^2 \over d\tau^2} r_{2n} & \simeq & s^2(r_{2n-2}-2r_{2n}+r_{2n+2})/4a^2 , 
\label{f25}
\end{eqnarray}
so that the equations of motion (\ref{f23}) transform to the system of discrete equations 
\begin{eqnarray}
{\cal V}'(r_{n+1})-2{\cal V}'(r_n)+ {\cal V}'(r_{n-1}) && \nonumber \\
+ \, \kappa(1-\bar{s}^2) (r_{n+2}-2r_n+r_{n-2})&=&0, 
\label{f26} 
 \end{eqnarray}
where $\bar{s}=s/2a\sqrt{\kappa}$ is the reduced value of the velocity.
It is convenient to look for the solution of the system of discrete equations numerically 
as a solution of the conditional minimum problem:
\begin{eqnarray}
{\cal F}=\sum_{n=1}^N\{ {\cal V}(r_{2n-1})+ {\cal V}(r_{2n})
+\frac12\kappa(1-\bar{s}^2) \nonumber \\
\times[(r_{2n-1}+r_{2n}-2)^2+(r_{2n}+r_{2n+1}-2)^2]\}
\nonumber \\
\rightarrow\min:\sum_{n=1}^N(r_{2n-1}+r_{2n})=2Na .
\label{f27}
\end{eqnarray}

The problem (\ref{f27}) has been solved numerically by the method of 
conjugated gradients described in Ref.~\cite{fr}. The initial point corresponding to the presence of the kink-antikink pair with the centers at the sites $N/4$ and $3N/4$ has been used as follows:
\begin{eqnarray}
r_{2n-1}&=&1-\delta, \nonumber \\
r_{2n}&=&1+\delta~~\mbox{for}~~n<N/4,~~n\ge 3N/4, \nonumber \\
r_{2n-1}&=&1+\delta,  \nonumber \\
r_{2n}&=&1-\delta~~\mbox{for}~~N/4\le n< 3N/4. 
\label{f28}
\end{eqnarray}
\begin{figure}[tbp]
\begin{center}
\includegraphics[angle=0, width=1\linewidth]{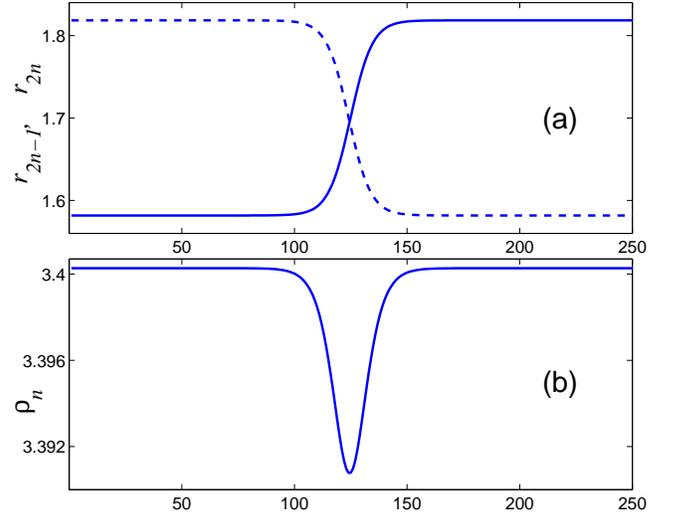}
\end{center}
\caption{\label{fig6}\protect\small (Color online)
Stationary topological soliton $(s=0)$ in stretched chain at $a=1.7$, $\beta=1$, $r_0=1$,
$\kappa=1$, $N=500$. Dependence of (a) relative displacements $r_{2n-1}$ ¨ $r_{2n}$ 
(solid and dashed lines) and (b) local compression $\rho_n=r_{2n-1}+r_{2n}$ on 
the number of the chain site are shown.
}
\end{figure}

Let $\{r_n\}_{n=1}^{2N}$ be a soliton solution of the problem 
(\ref{f27}). Then it is possible to find the soliton energy
\begin{eqnarray}
E=\frac12\sum_{n=1}^N\{ {\cal V}(r_{2n-1})+ {\cal V}(r_{2n})
+\frac12\kappa(1+\bar{s}^2) \nonumber \\
\times[(r_{2n-1}+r_{2n}-2)^2+(r_{2n}+r_{2n+1}-2)^2]\},
\label{f29}
\end{eqnarray}
and its diameter 
\begin{equation}
D=1+2\left[\sum_{n=1}^{N/2}(n+\frac12-\bar{n})p_n\right]^{1/2},
\label{29a}
\end{equation}
where the soliton center is given by
\begin{equation}
\bar{n}=\sum_{n=1}^{N/2}(n+\frac12)p_n,
\label{29b}
\end{equation}
and the sequence
\begin{equation}
p_n=(r_{2n+1}-r_{2n-1})/S,~~S=\sum_{n=1}^{N/2}(r_{2n+1}-r_{2n-1}),
\label{29c}
\end{equation}
determines the distribution of deformation along the chain.
\begin{figure}[tbp]
\begin{center}
\includegraphics[angle=0, width=1\linewidth]{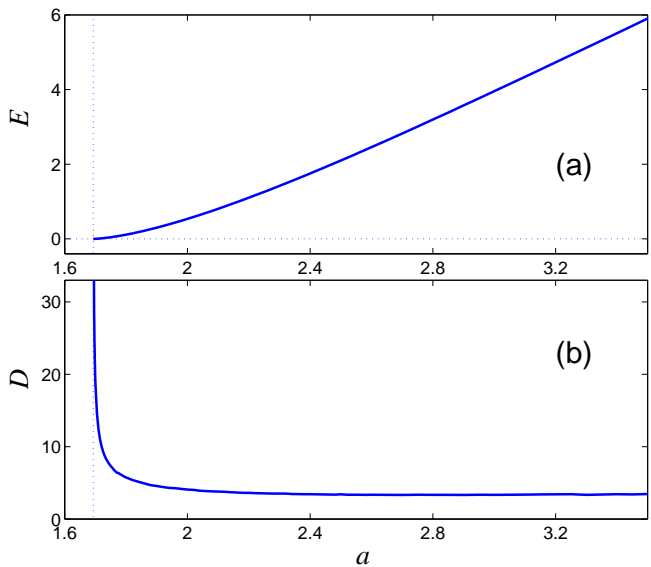}
\end{center}
\caption{\label{fig7}\protect\small (Color online)
Dependence of the energy of formation of soliton-antisoliton pair $E$ (a) and the diameter of topological soliton $D$ (b) on the value of lattice spacing $a$ of stretched cyclic chain consisting of $N=500$ sites ($\beta=1$, $r_0=1$, $\kappa=1$, and $a_0 = 1.693$).
}
\end{figure}

The shape of the topological soliton is presented in Fig.~\ref{fig6}. The panel (a) shows that the lengths 
of odd bonds $r_{2n-1}$ have the kink shape, whereas the lengths of even bonds $r_{2n}$ the 
antikink shape (and vice versa for the antikink). Next, as shown in the panel (b), the local compression of the chain takes place in the region of soliton localization.
\begin{figure}[tbp]
\begin{center}
\includegraphics[angle=0, width=1\linewidth]{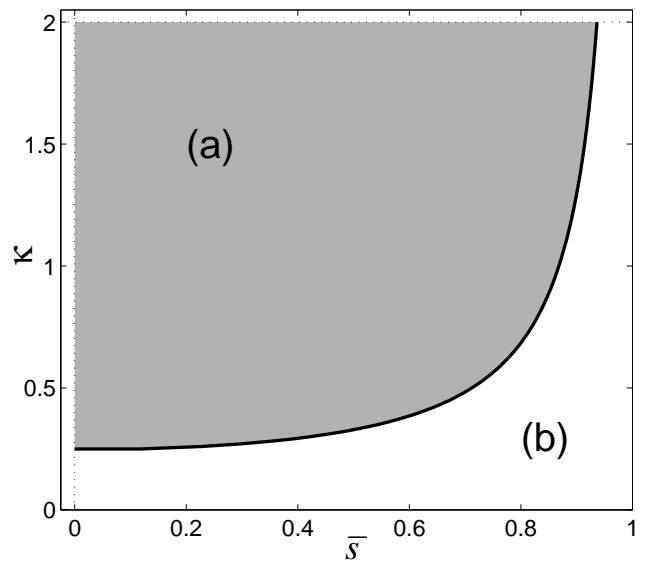}
\end{center}
\caption{\label{fig8}\protect\small
Region of existence of topological solitons in the space of parameters 
$\bar{s}$, $\kappa$ (a) 
and region of parameter values at which soliton solutions are absent (b). The line which splits these regions is given by (\ref{f30}).
 Lattice spacing is $a=1.7$ and $\beta=1$, $r_0=1$.
}
\end{figure}

The energy of formation of the kink-antikink pair can be defined as the difference 
$\Delta E=E-E_0$,
where $E$ is the energy of the stationary kink-antikink pair in the cyclic chain and $E_0$ is the 
energy of the ground state of the chain at a given lattice spacing $a>a_0$. 
As shown in Fig.~\ref{fig7}(a), the formation energy monotonically increases with the growth 
of the lattice spacing. Nearby the critical value of the lattice spacing $a_0$, the energy of formation becomes infinitesimal. When $a\rightarrow a_0+0$, the energy $\Delta E\rightarrow 0$ and 
the soliton diameter $D\rightarrow\infty$. With increasing $a$, the soliton diameter 
monotonically decreases down to the value $D=3.4$ [see Fig.~\ref{fig7}(b)].

Consider the dependence of the energy and the diameter of the soliton on its velocity. 
We choose the values: $\beta=1$, $r_0=1$, $a=1.7>a_0=1+\ln 2=1.693$ and 
$\kappa>\kappa_0=1/4$. 
It follows from Eq.~(\ref{f18}) that the reduced dimensionless velocity of sound is
\begin{equation}
\bar{s}_0=s_0/2a\sqrt{\kappa}=\sqrt{1+\frac{\kappa_-\kappa_+}{2\kappa(\kappa_-+\kappa_+)}}~,
\label{f30}
\end{equation}
where $\kappa_\pm=V''(a\pm\delta)$. The numerical solution of the problem (\ref{f27}) has shown 
that the system of discrete equations (\ref{f26}) has a soliton solution only in the subsonic 
region: $\bar{s}<\bar{s}_0$, $\kappa>\kappa_0$, being a typical situation for topological solitons (kinks). For our model, the region of the existence of topological 
solitons in the space of the parameters $\kappa$, $\bar{s}$ is shown in Fig.~\ref{fig8}.  
The region of the existence of solitons is separated from the region of their absence by curve 
(\ref{f30}) which determines the dependence of the reduced  velocity of sound $\bar{s}_0$ on the dimensionless stiffness $\kappa$.
\begin{figure}[tbp]
\begin{center}
\includegraphics[angle=0, width=1\linewidth]{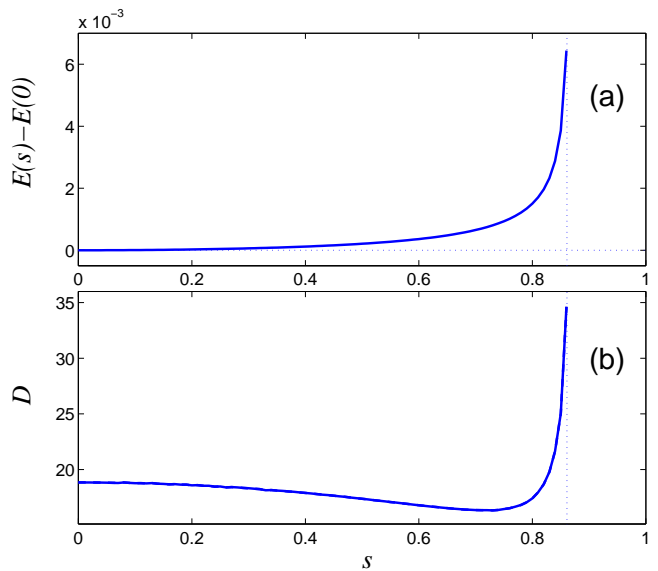}
\end{center}
\caption{\label{fig9}\protect\small (Color online)
Dependence of energy $E$ and diameter $D$ on reduced velocity 
$\bar{s}$ at $\beta=1$, $r_0=1$, $\kappa=1$, and $a=1.7$.
}
\end{figure}

The dependence of the energy and the diameter of the topological soliton on its velocity is given 
in Fig.~\ref{fig9}. As follows from this figure, the soliton energy monotonically increases with 
the growth the velocity. The energy tends to infinity at the velocity of long-wave acoustic 
phonons. The soliton diameter non-monotonically depends on its velocity. For small values the velocity increase results in negligible decrease of the diameter, which nearby the right edge of the velocity spectrum turns into the fast monotonic growth.

Consider the dynamics of a kink-antikink pair in a cyclic chain consisting of $N$ sites. 
To this end, we integrate the system of the equations of motion (\ref{f23}) with the initial conditions
\begin{eqnarray}
r_n(0)=r_n^0,~\mbox{for}~n=1,2, \ldots ,2N; \nonumber \\
r_{2n-1}'(0)=-s(r_{2n+1}^0-r_{2n-3}^0)/4a,\label{f31} \\
r_{2n}'(0)=-s(r_{2n+2}^0-r_{2n-2}^0)/4a,\nonumber \\
\mbox{for}~n=1,2, \ldots ,N, \nonumber
\end{eqnarray}
where $s$ is a soliton velocity and $\{r_n^0\}_{n=1}^{2N}$ is a soliton solution 
of the conditional minimum problem (\ref{f27}).
\begin{figure}[tbp]
\begin{center}
\includegraphics[angle=0, width=1\linewidth]{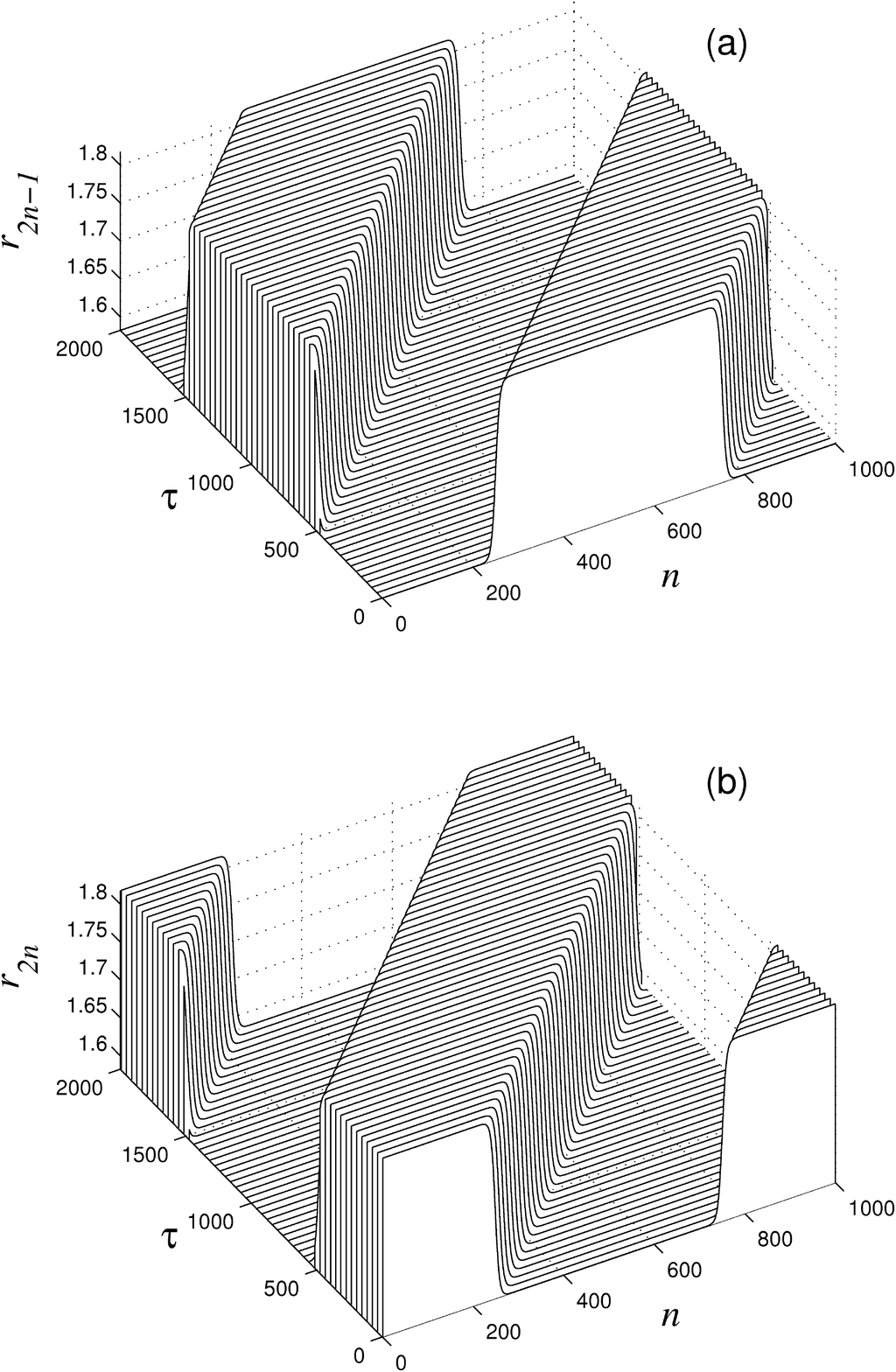}
\end{center}
\caption{\label{fig10}\protect\small
Uniform motion of kink-antikink pair in stretched chain. Dependence of chain distribution of 
odd $r_{2n-1}$ (a) and even $r_{2n}$ (b) bond lengths on time $\tau$. Parameter values: 
$\beta=1$, $r_0=1$, $\kappa=1$, $a=1.7$ (lattice spacing of stretched chain),  
$\bar{s}=s/2a\sqrt{\kappa}=0.5$ (reduced soliton velocity).
}
\end{figure}

The numerical integration of the system (\ref{f23}) with the initial conditions (\ref{f31}) has shown 
that the topological solitons in the stretched chain are dynamically stable for all admissible 
velocities $s<s_0$. As illustrated by Fig.~\ref{fig10}, the solitons move along the chain with a constant velocity without phonon radiation, completely retaining their initial shape.
\begin{figure}[tbp]
\begin{center}
\includegraphics[angle=0, width=1\linewidth]{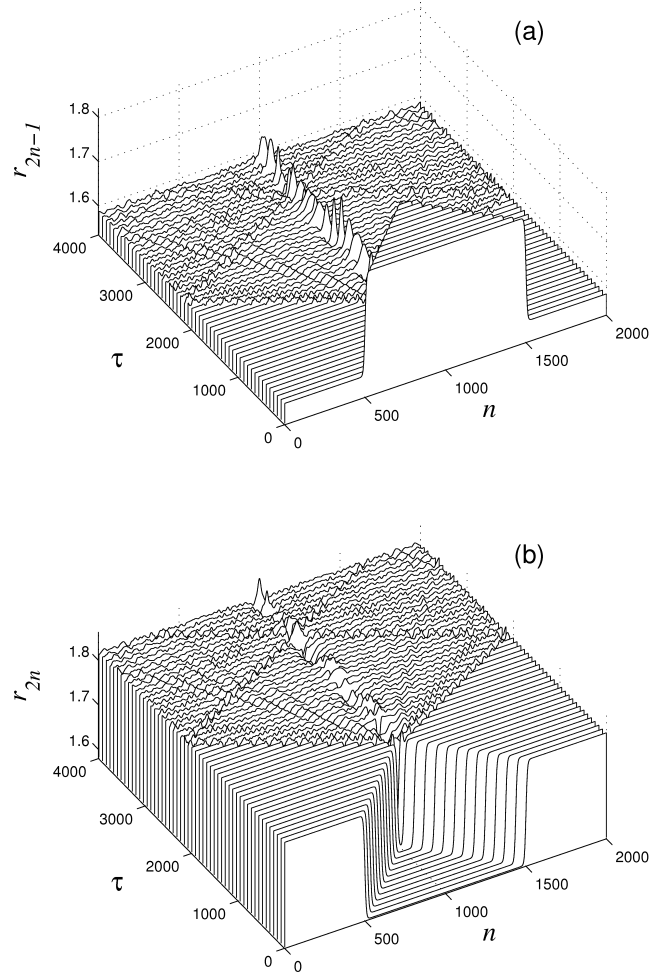}
\end{center}
\caption{\label{fig11}\protect\small
Annihilation of two topological solitons with opposite polarity under collision in stretched chain.
Dependence of the chain distribution odd $r_{2n-1}$ (a) and even $r_{2n}$ (b) bond lengths 
is shown. Parameter values: $\beta=1$, $r_0=1$, $\kappa=1$, $a=1.7$ (lattice spacing of stretched chain),
$\bar{s}=s/2a\sqrt{\kappa}=0.5$ (reduced soliton velocity).
}
\end{figure}

Consider now the interaction of the solitons with opposite polarity under their collision. 
To this end, we integrate the system (\ref{f23}) with the initial conditions
\begin{eqnarray}
r_n(0)=r_n^0,~\mbox{for}~n=1,2, \ldots ,2N; \nonumber \\
r_{2n-1}'(0)=-s(r_{2n+1}^0-r_{2n-3}^0)/4a, \nonumber \\
r_{2n}'(0)=-s(r_{2n+2}^0-r_{2n-2}^0)/4a,\nonumber \\
\mbox{for}~n=1,2, \ldots , N/2;\label{f32} \\
r_{2n-1}'(0)=s(r_{2n+1}^0-r_{2n-3}^0)/4a, \nonumber \\
r_{2n}'(0)=s(r_{2n+2}^0-r_{2n-2}^0)/4a,\nonumber \\
\mbox{for}~n=N/2+1,N/2+2, \ldots , N.\nonumber
\end{eqnarray}
The numerical integration has shown that this interaction is inelastic. Thus, at the velocity 
$\bar{s}=0.5$, the collision results in the annihilation of solitons with opposite polarity. 
The collision is accompanied by intensive phonon radiation and leads to the appearance 
a breather-like localized oscillation (see Fig.~\ref{fig11}).

For $\kappa<\kappa_0$ the chain can be found in the ground state of three types, depending on the lattice spacing $a$.  Under weak stretching 
$a\le a_1$ ($1<a_1<a_0$), the ground state is a uniformly stretched chain with equal bonds. 
At middle stretching  $a_1<a<a_2$ ($a_2>a_0$), a part of the chain is found in a weakly stretched homogeneous state with equal bonds and the spacing $a=a_1$, whereas the complementary part in a strongly
stretched state with alternating bonds and the spacing  $a=a_2$. For $a\ge a_2$ the whole chain is found in a homogeneous state with alternating bonds. 

\begin{figure}[tbp]
\begin{center}
\includegraphics[angle=0, width=1\linewidth]{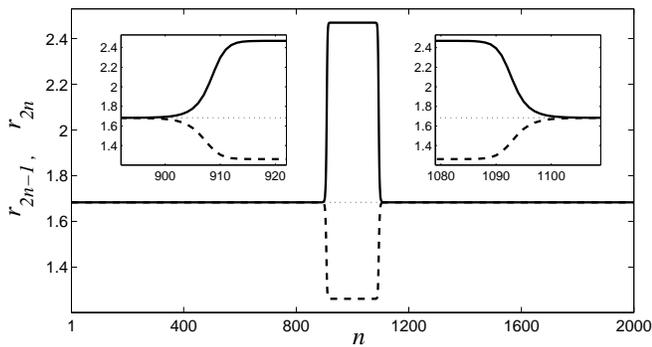}
\end{center}
\caption{\label{fig12}\protect\small
Non-uniform stationary state of stretched cyclic chain with $a>a_0$ and $\kappa<\kappa_0$ 
($\beta=1$, $r_0=1$, $a=1.7$, $\kappa=0.2$). Insets show 
that the edges of strongly stretched region 
of the chain have soliton shape (bond lengths $r_{2n-1}$, $r_{2n}$ smoothly depend on $n$).
}
\end{figure}

The numerical analysis confirms the existence of these three types of chain states. Thus, at $\beta=1$, $r_0=1$, $\kappa=0.2$, the critical values of the lattice spacing
are $a_0=1+\ln 2=1.6931$, $a_1=1.6833$, $a_2=1.8655$.
At $a=1.7$ the main part of the chain is found 
in the uniform state with equal bond lengths $a_1<a_0$, 
whereas the other part turns into the strongly stretched alternating state
with period $a_2>a_0$ (see Fig.~\ref{fig12}). 
This behavior of the chain under stretching can be explained by the 
non-convexity of the function $E(a)$ (see Sec. III).

\begin{figure}[tbp]
\begin{center}
\includegraphics[angle=0, width=1\linewidth]{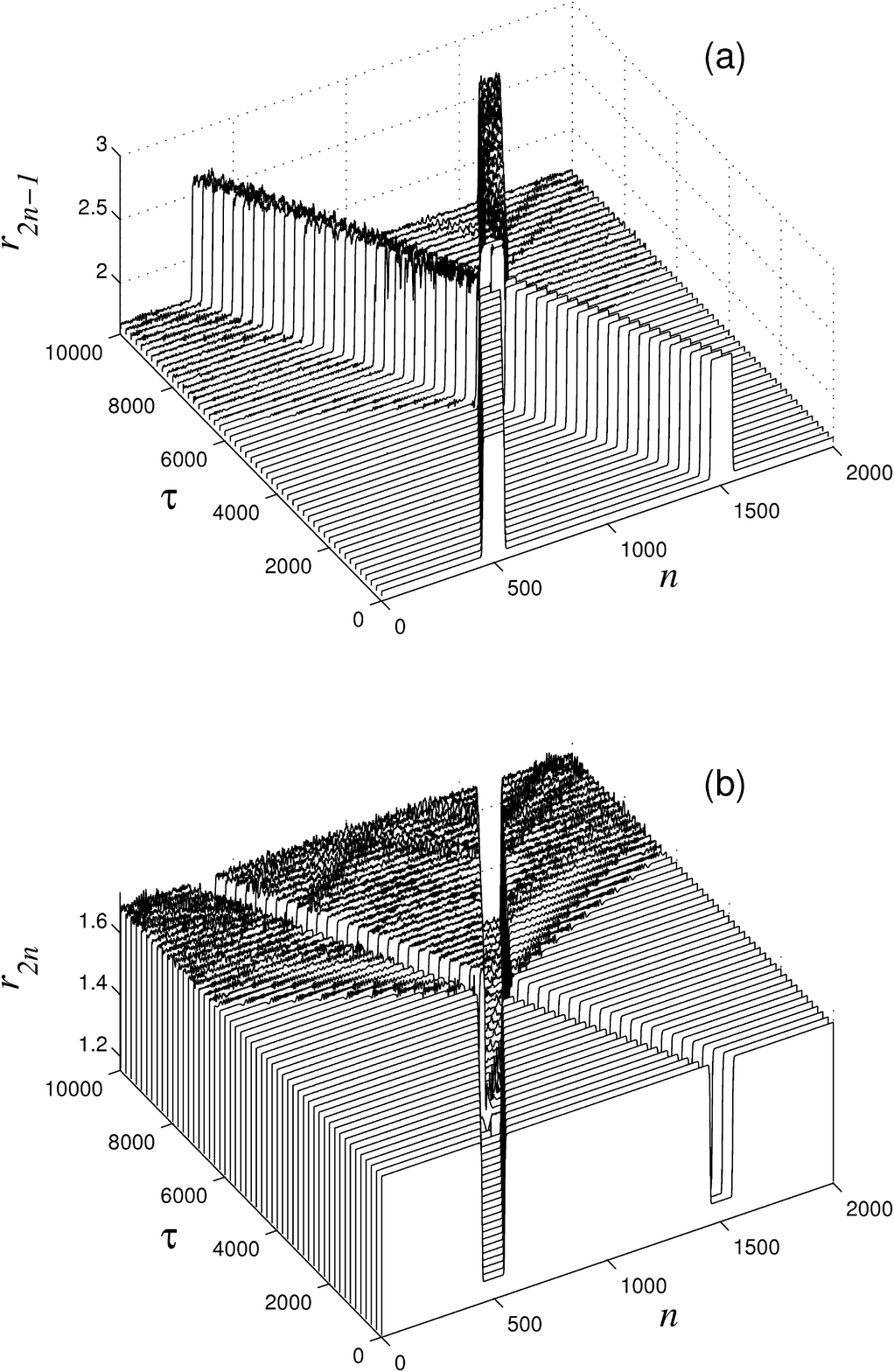}
\end{center}
\caption{\label{fig13}\protect\small
Inelastic collision of two strongly stretched regions in cyclic chain with $a>a_0$ and
$\kappa<\kappa_0$ ($\beta=1$, $r_0 =1$, $a=1.7$, $\kappa=0.2$, velocity is $\bar{s}=0.25$). 
Dependence of chain distribution of odd $r_{2n-1}$ (a) and even $r_{2n}$ (b) bond lengths 
on time $\tau$ is present.
}
\end{figure}

Figure \ref{fig12} also illustrates that the edges of the strongly stretched region of the chain with 
the alternating structure have the form of smooth stairs describing a smooth transition 
of the chain from the state with equal bonds to the state with alternating weakly and 
strongly stretched bonds. The numerical simulations have shown that the strongly stretched region 
can propagate along the chain with a subsonic velocity completely retaining its shape 
(see Fig.~\ref{fig13}). Therefore this transition region of the chain is a soliton describing the 
transition of only a part of the chain into the state with alternating bonds. 
As follows from this Figure, the collision of these solitons does not result in 
their destruction, but it is accompanied by phonon radiation.

\section{Examples of molecular chains where the existence 
of stretching solitons is possible}

We have studied the simplified one-dimensional lattice model. Nevertheless, this study 
allows us to define a family of molecular systems with quasi-one-dimensional structure in which 
the existence of topological solitons of stretching is possible.

The stability condition of alternating states of the stretched chain (\ref{f17}) imposes an 
important constraint. If we suppose that the stiffness of interaction of molecules is proportional 
to the energy of their interaction, then this condition can be treated as the requirement 
that the energy of interaction of the second neighbors in a quasi-one-dimensional molecular 
chain has to be no less than eighth part of the energy of interaction of the nearest neighbors. 
This condition of commensurability of the energies of interaction cannot obviously be realized 
in the chains where the nearest neighbors are coupled by strong valent bonds while the 
second neighbors by weak Van der Waals interactions.

Note that the condition (\ref{f17}) provides the stability of a stretched chain only with 
respect to its longitudinal deformations. In the three-dimensional space, the instability 
of a stretched chain can be caused by other (orientational, bending, twisting, etc.) deformations. 
Therefore the commensurability condition in this case is not sufficient.

Consider a zigzag-like chain of hydrogen bonds as the first example. Hydrogen fluorides HF, 
chlorides HCl, HBr, and HI (HX) at low temperatures have a crystalline structure formed by 
planar zigzag-like chains of hydrogen bonds \cite{s1,s2,s3,s4}. Consider an isolated 
hydrogen-bonded chain (HX$\cdots$)$_{\infty}$ consisting of two-atom molecules 
of fluoride HF and chloride HCl.

The interaction of two-atom polar molecules HX is usually described by the 12-6-1 potential \cite{s5}
\begin{equation}
 U= \sum_{i_1=1}^{3}\sum_{i_2=1}^{3} \frac{q_{i_1}q_{i_2}}{r_{i_1i_2}}+
 4\epsilon\left[\left(\frac{\sigma}{r}\right)^{12}
 -\left(\frac{\sigma}{r}\right)^6\right],
 \label{f33}
 \end{equation}
with the seven free parameters: two Lennard-Jones parameters $\epsilon$ and $\sigma$; 
three charges $q_1$, $q_2$, and $q_3$ ($q_1+q_2+q_3=0$), lying on the line of valent bonds, 
and three distances $r_1$, $r_2$, and $r_3$ which assign the charge positions. Here $r_{i_1i_2}$ is the distance between the charge $q_{i_1}$ of the first 
molecule HX and the charge $q_{i_2}$ of the second molecule given in terms of $r_1$, $r_2$, and $r_3$.

The values of the parameters for potential (\ref{f33}) can easily be found using 
the data of the crystalline structure of (HX)$_x$ and {\it ab initio} calculations of the 
dimer (HX)$_2$ \cite{s6}. We get 
\begin{eqnarray}  \nonumber
  & & q_1=-0.6397e,~q_2=0.6159e,~q_3=0.0238e,\\
 \nonumber
  & & r_1=0.25~\mbox{\AA},~~~r_2=0.9075~\mbox{\AA}
  ,~~~r_3=-1.575~\mbox{\AA},\\
 \label{d1} & & \epsilon=0.00798~\mbox{eV},~~~\sigma=2.837~\mbox{\AA}
\label{f34}
\end{eqnarray}
for hydrogen fluoride HF and
\begin{eqnarray}  \nonumber
  & & q_1=-0.3147e,~q_2=-0.7974e,~q_3=0.4827e,\\
 \nonumber
  & & r_1=1.296~\mbox{\AA},~~~r_2=-0.275~\mbox{\AA}
  ,~~~r_3=-0.838~\mbox{\AA},\\
 \label{d2} & & \epsilon=0.0298~\mbox{eV},~~~\sigma=3.602~\mbox{\AA}
 \label{f35}
\end{eqnarray}
for hydrogen chloride HCl, where $e$ is the electron charge.

In the plane of the zigzag-like chain ($\cdots$HX$\cdots$)$_{\infty}$, the position of each 
molecule HX is given by the coordinates $x$ and $y$ of the center of the heavy molecule X 
and the angle $\phi$ which shows the direction (orientation) of the molecule HX. 
The detailed description of the quantum-mechanical model of this zigzag-like structure is given 
in \cite{s7}. For hydrogen fluoride, the zigzag angle is $\alpha=119.5^\circ$, the longitudinal 
lattice spacing $l_x=2.167$~\AA, the distance between the nearest molecules 
$\rho_0=2.509$~\AA, 
the direction of each chain molecule differs from the zigzag line only by the angle 
$\varphi_0=01.21^\circ$. (The parameters for hydrogen chloride are $\alpha=93.6^\circ$, 
$l_x=2.692$~\AA, $\rho_0=3.694$~\AA, $\varphi_0=0.95^\circ$). In equilibrium, the energy 
of interaction of the nearest molecules is $E_1=0.2339$~eV and the energy of interaction 
of the second-neighboring molecules is $E_2=0.0312$~eV ($E_1=0.0890$~eV, $E_2=0.0165$~eV).

Here the commensurability condition (\ref{f17}) of the energy of interaction of the first and second 
neighbors is fulfilled. Thus, for the chain of molecules of hydrogen fluoride $E_2/E_1=0.133>1/8$ 
and for hydrogen chloride $E_2/E_1=0.185>1/8$. However, the analysis of spectrum behavior under 
the chain stretching has shown that the stability of a uniformly stretched state of the chain 
disappears before reaching the point of inflection of the effective potential of longitudinal 
stretching. In the case of HF, for the point of inflection the longitudinal lattice spacing 
of the zigzag is $a_0=2.67$~\AA, while the stability of the chain disappears already at the 
longitudinal lattice spacing $a_1=2.625$~\AA~(for HCl $a_0=3.60$~\AA, $a_1=3.48$~\AA). For the 
longitudinal spacing of the zigzag $a\ge a_1$, the chain becomes unstable with respect 
to the bending long-wave phonons, i.e., the bending instability of the chain takes place. 
Note that as regards the rest of (longitudinal and orientational) phonons, the chain keeps 
to be stable under the stretching $a>a_0$). Thus, the bending instability of zigzag-like 
chains does not admit the existence of the topological solitons of stretching. 
The maximum possible stretching of the chain can be defined as the ratio $a_1/l_x$. 
For the chain ($\cdots$HF$\cdots$)$_\infty$ the maximum stretching is 21\%,
while for the chain ($\cdots$HCl$\cdots$)$_\infty$ it is 29\%.

The similar bending instability of the chain is observed under stretching the trans-zigzag 
of the polyethylene macromolecule (---CH$_2$---)$_n$. The analysis of the linear dynamics 
of the planar zigzag of the chain within the model studied in Refs.~\cite{s7,s8} has shown that the strained 
chain is stable under stretching (the longitudinal lattice spacing of the zigzag) $a\le a_0$, 
where the critical value $a_0=1.745$~\AA~corresponds to the point of inflection of the 
effective potential of longitudinal stretching. However, under stretching $a>a_0$ the chain 
becomes unstable with respect to the bending oscillations of the chain. Here the maximum possible 
stretching of the chain is 37\% (in equilibrium the longitudinal lattice spacing of the zigzag 
is $l_x=1.276$~\AA). Here the bending instability is caused by the fact that the interaction 
of the second neighbors in the trans-zigzag occurs only because of the deformation of the 
valent angles CCC and it does not depend directly on the distance between them. 
The bending of the chain does not allow to break it without deforming the valent angles.

Thus, in a zigzag-like polyethylene macromolecule, the energy of interaction of the second 
neighbors is of the same order as the energy of interaction of the first neighbors, but the 
formation of bistable ground states is impossible due to the bending instability of the strongly 
stretched chain. For the absence of this instability the angles of a polymer chain have 
to possess a sufficiently strong nonvalent interaction of the second neighbors. The molecular 
groups of polyethylene CH$_2$ do not possess the interaction of this type. However, the 
radicals CHR with sufficiently long chains can provide this interaction. 
The polyolefine macromolecules
(---CH$_2$---CHR---)$_n$: polypropylene (R=C$_3$H$_7$), 
polystyrene containing benzol rings in radicals (R=C$_6$H$_5$)
and polyvinylcarbazole have the required stability structure. 
In these macromolecules with strongly interacting side radical groups R under the strong 
stretching of the chain, the topological solitons can exist. 

The existence of the topological 
solitons of stretching can be expected also in the DNA double helix. Here the conformational 
interaction of neighboring sites of the sugar-phosphate lattice plays the role of 
the first-neighbor coupling, whereas the stacking interaction of the neighboring purine 
and pyrimidine bases can be considered as the 
second-neighbor interaction. The experiments on stretching a single DNA 
molecule \cite{d1,d2,d3,d4} exhibit the presence of a specific constant region (plateau) 
in the force-stretching diagram  \cite{d5}. At strength about 
65 pN the non-typical behavior: the molecule becomes elongated at constant force up to 1.7 of its 
contour length. At further stretching the force again begins to grow. 
A similar behavior under stretching exhibit also $\alpha$-helices of protein \cite{d6}. 

Within our model this behavior will take place under a weak second-neighbor interaction, when $K<K_0$.
Here, in a certain interval of lengthening, the stretching occurs according to the two-phase scenario,
when one part of the chain is found in a weakly and the other one in a strongly stretched state. 
In the force-stretching diagram a constant region (plateau) appears due to the stretching 
of the chain because of increasing only the portion of its strongly stretched part. Obviously, 
here the topological solitons describing the transition from the weakly to the strongly stretched 
phases of the chain have to exist. 

\section{Conclusions}

The study carried out in this paper shows under the stretching of molecular chains the 
conformational changes of these chains can occur that result in different ground states.  
The transition regions between these states can be described as topological solitons. 
In the simplest model with the nearest-neighbor interaction of the Morse-like type 
and the second-neighbor harmonic interaction, it is shown that under the chain stretching 
the ground state is realized as a regular configuration with alternating bonds 
(``long-short''). In this case, the chain can be found in two degenerate ground states admitting 
the existence of topological solitons that describe the chain transition from the state 
``short-long bond'' into the state ``long-short bond''. This situation is possible for the 
molecular chains with sufficiently strong interaction of the second neighbors. 
With weak interaction, the chain stretching leads to the appearance of one region with weakly 
and the other one with strongly stretched bonds. As a result of this non-uniform stretching, 
the presence of a broad plateau in the force-stretching diagram of DNA double helix 
and protein $\alpha$-helix can be explained. The boundary between the weakly 
and strongly stretched phases of the chain can also be described as a topological soliton.       

Finally, it should be noticed that the models with the second-neighbor coupling being responsible 
for the stabilization of homogeneous bistable ground states have been studied earlier 
\cite{s7,zps91,k,zps00,kzcz}. However, in these (diatomic) models, the effect of switching 
or controlling bistability by external forcing has not been considered. The bistability here 
has been attained intrinsically due to the repulsive interaction in the heavy-ion sublattice. 

\begin{center}
{\bf  ACKNOWLEDGMENTS}
\end{center} 

A.V.Z acknowledges the partial financial support
from the Ukrainian State Grant for Fundamental Research. 
Both of us (A.V.S. and A.V.Z.) would also like to express
his gratitude to the MIDIT Center and Department of Informatics and Department
of Physics of the Technical University of Denmark
for partial financial support and hospitality.

\end{document}